# Temperature Profile for Ballistic and Diffusive Phonon Transport in a Suspended Membrane with a Radially Symmetric Heat Source

J.T. Karvonen[1,2], T. Kühn[2], and I.J. Maasilta[2,*]

[1]*I. Physikalisches Institut IA, RWTH Aachen University, 52056 Aachen, Germany*
[2]*Nanoscience Center, Department of Physics, University of Jyväskylä, FI-40014, Finland*

We have calculated the temperature profiles for phonon heat transport in a suspended membrane with a radially symmetric heat source, in the two extreme cases of either fully ballistic or fully diffusive transport. Theoretical results confirm that it is possible to distinguish these two transport mechanisms from the radial temperature profiles alone. Models are also compared to experimental data measured with 40 nm thick, free standing silicon nitride membranes below 1 K by using tunnel junction (SINIS) thermometers. The measured temperature profile is in quantitative agreement with the ballistic model for distances below 50 µm, but not above.

## I.   INTRODUCTION

The nature of the thermal transport mechanism is one of the important questions when low temperature applications of thermal devices are designed. Suspended silicon nitride ($SiN_x$) membranes are widely used platforms for many low temperature applications due to their efficient thermal isolation from environment and fairly straightforward fabrication processes. They are widely used in bolometric radiation detectors from sub-mm to gamma rays, for example [1]. Traditionally, transport mechanisms in these membranes have been studied by measuring the temperature dependence and the absolute value of the heat transport [2-5], but the conclusions have been contradictory. It is thus safe to say that phonon thermal transport in thin SiN membranes is not yet deeply understood.

Here, we discuss another approach to solve this problem, by studying the temperature profile, *T(r)*, of the membrane. This entails not only measurements of how the temperature of the membrane rises with applied heat at one location, but at several locations as a function of the distance from the heater. This technique will give us additional information about the nature of phonon transport in the membranes. Specifically, we first calculate the temperature profile in the two extreme cases of fully ballistic and fully diffusive heat transport, in the simple geometry of a radially symmetric heater and thermometer on top of the membrane. The results clearly show that it is possible to distinguish these two transport mechanisms purely from the shape of the *T(r)* profile. Finally, we compare the results to experimental data measured for a 40 nm thick free-standing $SiN_x$ membrane below 1 K. The measured data is qualitatively more in agreement with the profile from the ballistic model, although it starts to deviate form it at larger distances.

## II.   THEORETICAL CONSIDERATIONS

Calculations are simplified a lot if one considers the highly symmetric case of a large membrane, with a small circularly symmetric heater in the middle of the membrane, and a narrow thermometer wire in the shape of a circular arc at a distance *r* from the center of the heater. The schematic of the geometry is presented in Figure 1. The fact that heat can only flow along the membrane restricts the problem first to two dimensions, and the radial symmetry of the heater reduces in to one dimension. Thus, temperature will only depend on the radial coordinate *r*. In principle, full radial symmetry would require the membrane edge (where it connects to the bulk substrate) to be circular in shape, also. However, if this membrane edge is far away from the heater and the thermometer, the actual shape of the membrane edge plays no role, as the corrections to the temperature profile would appear only near the edge. This consideration is relevant in practice, because even though we can fabricate a circular heater using lithography, the membrane edge is a square due to the crystallographic etching process used to suspend the membrane [6].

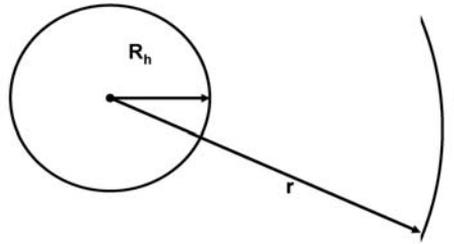

FIG.1: Schematic of the studied sample geometry. A circular heater with radius $R_h$ emits phonons, which can be detected by a thermometer of length *l* distance *r* away from the center of the heater.

First, we consider the case of ballistic phonon transport, where no scattering takes place in the bulk of the membrane, and all surface scattering is specular. This limit is the extreme case of a more general class of radiative transport models, where some or all of the surface scattering can be diffusive [7]. In these models one uses the analogy between phonon and photon radiation, where the radiation in vacuum is in dynamic equilibrium with emitting and absorbing surfaces at given temperatures. In the case of ballistic phonon transport, the surfaces are the metallic heater and thermometer, which interact with the phonons because of electron-phonon coupling, and the analog of vacuum is the SiN membrane itself. One immediate important result is then obvious without any explicit calculations: the geometry (size) of the heater and the thermometer will influence the temperature profile. In other words, one can define and measure the temperature of the thermometer in equilibrium with the phonon radiation, but that temperature depends on the size of the thermometer and heater in the general case. Another corollary of this result is that if one places two or more thermometers on the membrane, they influence each other so that the measured temperatures change. This was actually observed experimentally [6].

In the following, we also simplify the calculations by assuming that the substrate is at zero K. As our experiments are performed usually at very low substrate temperatures *T ~ 50* mK, the error introduced is small if the measured phonon temperatures are higher than that. Finally, if the membrane edge is fully absorbing (black body at *T=0* K), we can ignore it altogether in the calculation. Within these assumptions, in steady state, the net radiative powers emitted by the heater, $P_h$, and thermometer, $P_t$, are

$$P_h = P_{heat} + \alpha P_t$$
$$P_t = \beta P_h, \tag{1}$$

where $P_{heat}$ is the power input from the electrical circuit into the heater, $\alpha = R_h/2r$ and $\beta = l/2\pi r$, where $R_h$ is the radius of the heater, $r$ the distance between the center of the heater and the thermometer and $l$ the length of the thermometer arc. The factors $\alpha$ and $\beta$ can be derived from geometry: $\beta P_h$ is the fraction of power emitted by the heater that is intercepted by the thermometer, and $\alpha P_t$ is the fraction of power emitted by the thermometer that is intercepted by the heater. In the above, we also assumed that the thermometer doesn't intercept any of its own radiation, which is a good approximation if $l << r$. From equations (1), we can then solve for $P_t$

$$P_t = \frac{l}{2\pi r}\left(1 - \frac{lR_h}{4\pi r^2}\right)^{-1} P_{heat} \equiv P_{rad}, \tag{2}$$

which is equal to the emitted power of the thermometer from a phonon radiative transfer model, $P_{rad}$. When an effective model $P_{rad} = aT^n$ with a constant $a$ is used, we finally obtain the distance dependence in the ballistic limit

$$T = \left[\frac{Bl}{2\pi r}\left(1 - \frac{lR_h}{4\pi r^2}\right)^{-1}\right]^{(1/n)}, \tag{3}$$

where $B = P_{heat}/a$. The dominant dependence at large distances $2\pi r > l$ is then simply $T \sim 1/r^{(1/n)}$, when the finite size correction (term in parentheses) can be ignored. The constants $a$ and $n$ can be thought of as fitting parameters from a fit to $T$ vs. $P_{heat}$ data at constant $r$. Theory does, however, predict for 3D phonons in the isotropic case that $n = 4$ and $a = 2ld\sigma$, where $d$ is the membrane thickness, $\sigma$ the phononic Stefan-Boltzmann constant

$$\sigma = \frac{\pi^2 k_B^4}{120\hbar^3}\left(\frac{2}{c_t^2} + \frac{1}{c_l^2}\right), \tag{4}$$

and where $c_t$ and $c_l$ are the transverse and longitudinal speeds of sound. Thus, we obtain in the limit where the thermometer is small ($l << R_h$)

$$T(r) = \left(\frac{P_{heat}}{2\pi\sigma dr}\right)^{1/4}. \tag{5}$$

Interestingly, this temperature profile has a universal shape, and its magnitude is scaled only by the material parameters through $\sigma$ and the thickness of the membrane $d$.

In the diffusive case, one can define, as usual, a local phonon temperature without the explicit consideration of the thermometer geometry. The radial heat flow is written as

$$P = -S(r)\kappa(T)\frac{dT}{dr}, \tag{6}$$

where $S(r) = 2\pi rd$ is the cross sectional area for the heat transport for the studied geometry at a distance $r$ from the center and $d$ is again the thickness of the membrane. Now we assume for the thermal conductivity $\kappa(T) = bT^m$, where $b$ is a constant, and from equation (6) we get

$$-\int_{r_0}^{r}\frac{P}{2\pi rd}dr = \int_{T_0}^{T}bT^m dT \tag{7}$$

After noting that the total power cannot depend on $r$, and integrating, equation (7) gives

$$\frac{P}{2\pi d}\ln\left(\frac{r_0}{r}\right) = \frac{b}{m+1}\left(T^{m+1} - T_0^{m+1}\right), \qquad (8)$$

where a point $(r_0, T_0)$ is known. From equation (8) we can solve the temperature profile in the diffusive case:

$$T = \left[T_0^{m+1} + \frac{(m+1)P}{2\pi bd}\ln\left(\frac{r_0}{r}\right)\right]^{1/(m+1)}, \qquad (9)$$

where the power $P$ must be equal to the power input from the electrical circuit, $P=P_{heat}$ and is therefore known. In our experimental situation, the only fixed boundary condition is at the membrane edge, where the temperature must equal the substrate temperature. Thus, in Eq. (9) $r_0$ is the membrane edge, and with the same approximation as before that substrate temperature equals zero $T_0=0$ K. The temperature of the heater is not fixed, but depends on the input power. We thus note a significant difference between the diffusive and ballistic cases: The diffusive model requires the presence of the edge of the membrane, whereas the ballistic conductance can be considered for an infinite membrane also.

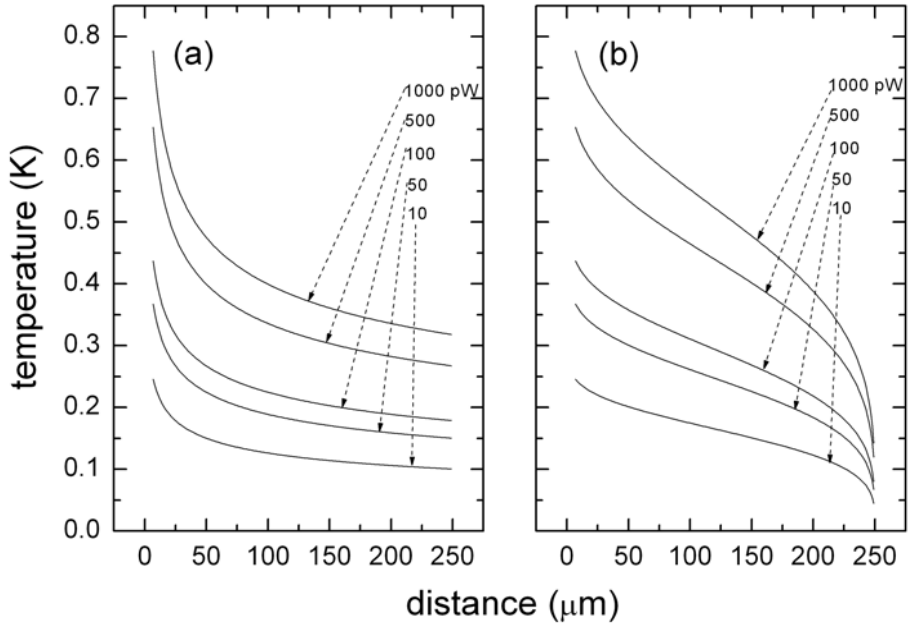

FIG.2: Theoretical temperature profiles for (a) ballistic [Eq. (5)] and (b) diffusive [Eq. (9)] phonon transport in membranes for a radially symmetric geometry. Different curves correspond to varying input powers in sequence $P= 10, 50, 100, 500, 1000$ pW.

In Figure 2 we compare the temperature profiles calculated either from the ballistic or diffusive models, Eqs. (5) and (9). In the diffusive case we have chosen the constant in thermal conductivity, $b$, in such a way that the temperatures at the heater location are equal for both models for the same input power $P_{heat}$. The plots are calculated using the speeds of sound for

SiN $c_t$ = 6200 m/s and $c_l$ = 10 300 m/s and a membrane thickness $d$ = 200 nm. The heater is located at $r$ = 7 μm, and the membrane edge at $r$ = 250 μm. Furthemore, we choose $n = m+1 = 4$, so that in both cases $P \sim T^4$. Physically, m = 3 corresponds to diffuse surface scattering, which is expected to be the dominant scattering mechanism at low temperatures [8].

From Fig. 2 we clearly observe that the temperature profiles have quite different shapes for the two cases, and that the shape does not vary with input power. The ballistic case [Fig. 2(a)] has a steep drop close to the heater, and then a slow decay at large distances, whereas the diffusive profile [Fig. 2(b)] is almost linear, except for a steeper initial drop near the heater, and then a strong drop near then membrane edge. At the membrane edge the two cases have a completely different behavior, as in the ballistic case the edge plays no role. The profiles should also be compared to the well known 1D case [7], where the temperature profile is constant or linear in the ballistic and diffusive cases, respectively.

## III. EXPERIMENT

The measured samples have a ~40 nm thick, suspended low-stress $SiN_x$ membrane (size ~550x550 μm$^2$), which were anisotropically etched from double-side, LPCVD nitridized (100) silicon wafers in an aqueous KOH solution. At the center of the membrane, we fabricate a circular copper wire heater of radius 7 μm, width ~200nm and thickness 30 nm, which is directly connected to superconducting Nb leads at both ends of the heater wire, forming SN junctions. A symmetric normal metal-insulator-superconductor (NIS) tunnel junction pair made from Cu/AlO$_x$/Al is also fabricated at the distance $r$ from the center to measure the local phonon temperature of the membrane. Several samples were made with varying distances between the heater and the thermometer, keeping the heater and thermometer sizes constant, to study the temperature profile. Further details of the sample fabrication process and geometry are presented in Ref. [6].

In the measurement, we apply a slowly ramping DC voltage into the heater wire and measure the input Joule heating power $P=IV$ by a four probe configuration. Due to the Andreev reflection in the SN contacts, all the input power is dissipated uniformly only in the Cu wire, causing therefore a radially symmetric phonon emission power (via electron-phonon coupling) into the membrane. Simultaneously, the current biased SINIS tunnel junction thermometer measures the response in the local phonon temperature $T_p$ at distance $r$. A more detailed description of the measurement technique and tunnel junction thermometry is presented in references [6,9].

In Figure 3 we present measured data (black dots) for one input heating power value $P$=0.2 nW for nine different distances. The measured temperature profiles look very similar for other values of $P$, as expected. At this level of heating, the measured temperatures are high enough that the membrane phonons are expected to be in the 3D limit [9]. For comparison, we have also plotted the theoretical models in the ballistic and diffusive cases from Equations (3) and (9). For the diffusive model, we used the temperature exponent $m$ = 3, and fixed the fitting parameter $b$ such that the measured data agrees with the model at two points $(r_0,T_0)$=(12.2 μm, 802 mK) and $(r_1, T_1)$=(250 μm, 50 mK) (the membrane edge). For the ballistic model, we used $n$ = 4 and the fitting constant is $B$=7.8x10$^{11}$ K$^4$.

Clearly the measured data agrees qualitatively better with the shape of the ballistic model, especially at short distances $r < 50$ μm. The diffusive model would predict a temperature profile almost linear and much higher in absolute value than the measured data. After the strong decrease in temperature, the measured temperatures form a plateau up to $r \sim 150$ μm, where the temperature hardly decreases at all with distance. Although the ballistic model predicts a flat temperature profile at large distances, the absolute values of the measured temperatures are much higher than what is expected from the simplest ballistic theory, Eq. (3). This is not fully understood at the moment, but might be caused by a fraction of the surface scattering events being diffusive, increasing the thermal resistance. Also, at the last point, $r=180$ μm, the measured temperature drops again quite significantly compared to the previous points. This is most likely due to the effect of the membrane edge, but understanding of this behavior is currently lacking, also.

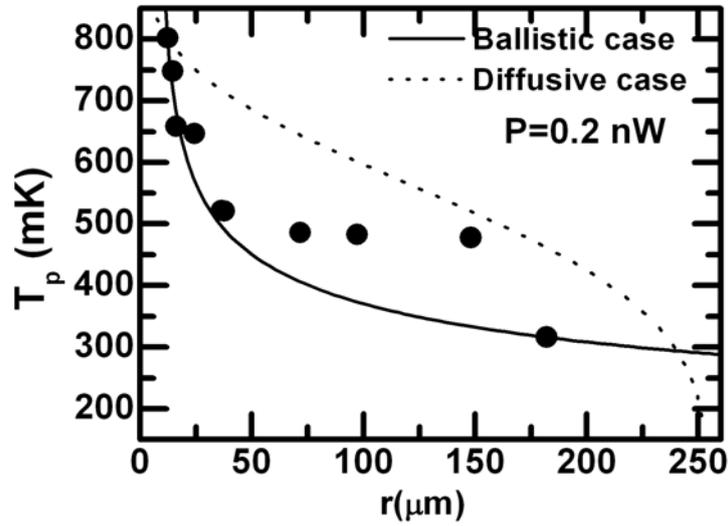

FIG.3: The local phonon temperature $T_p$ versus the distance from the center $r$ at constant heating power $P=0.2$ nW. Black dots: measured data. Black solid line: the ballistic model from Eq. (3). Black dotted line: the diffusive model from Eq. (9).

## IV. CONCLUSIONS

We have studied theoretically and experimentally the temperature profile of thermal transport in thin suspended $SiN_x$ membranes at sub-Kelvin temperatures. The radial temperature profile has been calculated for a circular symmetry, for either fully ballistic or fully diffusive transport. It is clear that these two transport mechanism have very different temperature profiles even in the case of localized heating in membranes with 2D heat flow, much like in the better known case of 1D heat flow [7]. Theoretical models were also compared to the measured data, which agrees qualitatively with the ballistic model. However, full agreement with the simplest ballistic theory and experiment was not achieved. This is, in hindsight, not a big surprise, as we only considered the simplified limit where all the surface scattering is specular. By introducing a more refined model with some portion of diffusive surface scattering, better agreement with experiment might be possible.

**ACKNOWLEDGEMENT**

This work is supported by the Academy of Finland projects No. 118665 and 118231. T. K. thanks Emil Aaltonen Foundation for financial support.